# Charge Density Wave order and electron-phonon coupling in ternary superconductor Bi$_2$Rh$_3$Se$_2$

Zi-Teng Liu,[1] Chen Zhang,[1] Qi-Yi Wu,[1] Hao Liu,[1] Bo Chen,[1] Zhi-Bo Yin,[1] Sheng-Tao Cui,[2] Zhe Sun,[2] Shuang-Xing Zhu,[1] Jiao-Jiao Song,[1] Yin-Zou Zhao,[1] Hong-Yi Zhang,[1] Xue-Qing Ye,[1] Fan-Ying Wu,[1] Shu-Yu Liu,[1] Xiao-Fang Tang,[3] Ya-Hua Yuan,[1] Yun-Peng Wang,[1] Jun He,[1] Hai-Yun Liu,[4] Yu-Xia Duan,[1] and Jian-Qiao Meng[1, *]

[1]*School of Physics and Electronics, Central South University, Changsha 410083, Hunan, China*
[2]*National Synchrotron Radiation Laboratory, University of Science and Technology of China, Hefei 230029, Anhui, China*
[3]*Department of Physics and Electronic Science, Hunan University of Science and Technology, Xiangtan 411201, Hunan, China*
[4]*Beijing Academy of Quantum Information Sciences, Beijing 100085, China*
(Dated: Sunday 6$^{th}$ March, 2022)

The newly discovered ternary chalcogenide superconductor Bi$_2$Rh$_3$Se$_2$ has attracted growing attention, which provides an opportunity to explore the interplay between charge density wave (CDW) order and superconductivity. However, whether the phase transition at 240 K can be attributed to CDW formation remains controversial. To help resolve the debate, we study the electronic structure study of Bi$_2$Rh$_3$Se$_2$ by angle-resolved photoemission spectroscopy experiments, with emphasis on the nature of its high-temperature phase transition at 240 K. Our measurements demonstrate that the phase transition at 240 K is a second-order CDW phase transition. Our results reveal (i) a 2 × 2 CDW order in Bi$_2$Rh$_3$Se$_2$, accompanied by the reconstruction of electronic structure, such as band folding, band splitting, and opening of CDW gaps at and away from Fermi level; (ii) the existence of electron-phonon coupling, which is manifested as an obvious kink and peak-dip-hump structure in dispersion; and (iii) the appearance of a flat band. Our observations thus enable us to shed light on the nature of the CDW order and its interplay with superconductivity in Bi$_2$Rh$_3$Se$_2$.

Charge density wave (CDW) and superconductivity are two very important and closely linked charge orders in solid. According to BCS superconducting theory [1] and CDW formation mechanism [2, 3], both superconductivity and CDW state are the result of electron-phonon (*e-ph*) coupling, and CDW state usually competes with the superconducting state. After over 35 years of intensive research, the underlying mechanism of high-$T_c$ superconductivity remains a central mystery in the physics of condensed matter [4, 5]. More and more evidence shows that there is charge order or the CDW in high-$T_c$ cuprate superconductors [6–11]. However, the relationship between CDW and high-$T_c$ superconductivity is difficult to determine whether they coexist or compete. Even in BCS superconductors, CDW does not just compete with superconductivity [12].

Bi$_2$Rh$_3$Se$_2$, which coexists structural phase transition and superconductivity, provides a good opportunity to study CDW and its interaction with superconductivity [13–15]. Bi$_2$Rh$_3$Se$_2$ undergoes a superconducting transition at $T_c \sim 0.7$ K along with a phase transition at $T_s \sim 240$ K [13]. Both $T_c$ and $T_s$ can be tuned by pressure or element substitution. The $T_s$ of Bi$_2$Rh$_3$Se$_2$ increases linearly with pressure and can rise above 300 K at 22.23 kbar [15]. Replacing Rh with Pd and Ni, the $T_c$s for Bi$_2$Pd$_3$Se$_2$ [16] and Bi$_2$Ni$_3$Se$_2$ [17] are 0.73 and 0.96 K, respectively, but no high-temperature phase transition is found in both. While replacing Se with S, the $T_s$ of Bi$_2$Rh$_3$S$_2$ can be reduced to 165 K, and there is no superconductivity below 0.5 K [18]. By further increasing the Rh content, isostructural compound Bi$_2$Rh$_{3.5}$S$_2$ has higher bulk superconductivity ($T_c \sim 1.7$ K), but no structural phase transition is observed at high temperatures [18]. Although many studies have been carried out on Bi$_2$Rh$_3$Se$_2$, whether the phase transition occurs at $\sim$ 240 K is a CDW order is still controversial [13–15]. Sakamoto *et al*. suggested that the phase transition was a CDW order by transport and low-temperature X-ray measurements [13]. Lin *et al*.'s optical spectroscopy measurement found that an energy gap was formed with associated spectral changes only at low energy, thereby thinking it is a CDW phase transition [14]. However, by pressure and selected-area electron diffraction studies, Chen et al. came to a different conclusion, that is, it is not a CDW transition, but a pure structural phase transition accompanied by symmetry reduction [15]. Understanding the electronic structure resulting from phase transition is crucial to revealing the nature of phase transition and its possible relation to superconductivity. Angle-resolved photoelectron spectroscopy (ARPES) is a powerful tool to study electronic structure, but there is no relevant experiment at present.

In light of the controversy about the origin of high-temperature structural phase transition and its possible connection with superconductivity, we report in this paper detailed temperature dependence of electronic structures in Bi$_2$Rh$_3$Se$_2$ by ARPES measurements. We reveal the existence of a 2 × 2 CDW order with a transition temperature of 244 ± 4 K. We observed the CDW-induced band folding, band splitting, and the opening of the CDW gaps at and away from the Fermi level. In particular, the signatures of *e-ph* coupling, kink and peak-dip-hump structure were observed at $\sim$ 52 meV.

Data presented in Figs. 1(c), 2, 3, 4, and 5 were taken at the BL13U beamline in National Synchrotron Radiation Laboratory (NSRL, Hefei), which was equipped with a Scienta DA30 analyzer. The samples were cleaved at 30 K under a vacuum better than 1 × 10$^{-11}$ mbar. Data presented in Fig. 1(b) was obtained at our home experimental system equipped with the Scienta R4000 electron energy analyzer. We use Helium I resonance line as light source with a photon energy of $h\nu$ = 21.218 eV. The samples were cleaved at 5 K under a vacuum better than 1 × 10$^{-11}$ mabr. All data collection was performed on the freshly cleaved sample surface. High-quality Bi$_2$Rh$_3$Se$_2$ single crystals were grown by the self-flux

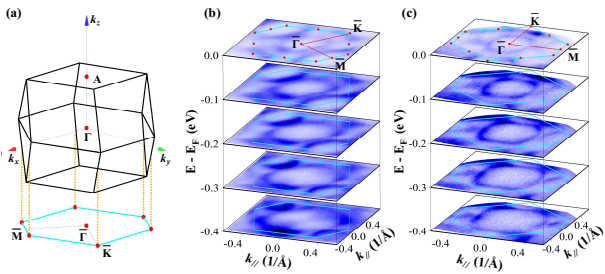

FIG. 1. **BZ and constant energy contours of $Bi_2Rh_3Se_2$.** (a) The bulk BZ and the projected surface BZ for (001) surface with high-symmetry momentum points marked (red dots). (b) and (c) Fermi surface maps of $Bi_2Rh_3Se_2$ taken at photon energies of 21.218 eV and 30 eV respectively, and temperatures of 250 K and 270 K respectively. Constant-energy contours are at Fermi energy and at 100, 200, 300, and 400 meV below $E_F$, integrated over 10 meV.

method.

In the normal state, $Bi_2Rh_3Se_2$ crystallizes in a parkerite-type structure with the $C12/m1$ space group [13]. The bulk Brillouin zone (BZ) of $Bi_2Rh_3Se_2$, as well as its projection on (001) surface BZ, are shown in Fig. 1(a). The bulk BZ is a rhombic dodecahedron, and the surface BZ is nearly hexagonal. Figures 1(b) and 1(c) show the Fermi surfaces of $Bi_2Rh_3Se_2$ measured with 21.218 eV (helium lamp) and 30 eV photons, respectively. The Fermi surface is composed of large nearly elliptical pockets. With the increase in binding energy, the Fermi pocket centered on $\overline{\Gamma}$ gradually becomes larger, indicating that it is a hole pocket. The Fermi pockets located at the BZ boundary become smaller with the increase of binding energy, which is proven to be electron pockets.

Figures 2(a1) and 2(b1) show the broad-range energy dispersion maps along the $\overline{K}$-$\overline{\Gamma}$-$\overline{K}$ direction above (260 K) and below (30 K) $T_s$, respectively. It can be seen that only the band structures near the Fermi level change significantly. While the band structures at higher binding energy ($< -1$ eV) remain unchanged after phase transition. Figures 2(a2) and 2(b2), the corresponding second derivative images of Figs. 2(a1) and 2(b1), show clearer energy band structure, and the stability of deep energy band dispersion can be seen intuitively. If $Bi_2Rh_3Se_2$ undergoes a primary structure phase transition at 240 K, its band structures, including the structures at higher binding energy, should be strongly modified [19–22], but this is not the case here. Changes in the low-energy band structure, such as the disconnection of original continuous bands due to the opening of the energy gaps, indicate that $Bi_2Rh_3Se_2$ undergoes a CDW phase transition rather than a pure structural phase transition.

To study the possible CDW phase transition, the temperature-dependent low-energy electronic structure was measured. Figures 3(a1)-3(d1) show the original band structures measured along the high-symmetry $\overline{M}$-$\overline{\Gamma}$-$\overline{M}$ direction at labeled photon energies and temperatures (see Figs. S1 and S2 of the Supplemental Material [24] for more details). By tracking the temperature dependence of the spectral intensity of the region where the replica bands appear, the phase transition temperature $T_{CDW} = 244 \pm 4$ K is obtained, which is consistent with resistivity measurement [13] (see Fig. S2 and S3 of the Supplemental Material [24] for more details). Figures. 3(a2)-3(d2) and Figs. 3(a3)-3(d3) display the second derivative images of Figs. 3(a1)-3(d1) with respect to momentum and energy, respectively. In the normal state, it is clearly shown that multiple linear dispersions cross the Fermi level, indicating that it is a metallic state. The strongest band approaches the $\overline{\Gamma}$ point and forms a hole pocket [Figs. 3(a) and 3(c)], and two weaker bands form two electron pockets around the $\overline{M}$ points [Fig. 3(a)]. A very weak band that probably does not cross the Fermi level is also observed at the $\overline{\Gamma}$ point [Figs. 3(a) and 3(c)]. As shown in Figs. 3(b) and 3(d), these light bands remain in the CDW state and become sharper due to a reduced thermal broadening, especially the one that does not cross the Fermi level. Importantly, the electronic structure reconstruction is directly observed in the measured [Figs. 3(b1) and 3(d1)] and second derivative images [Figs. 3(b2), 3(b3), 3(d2), and 3(d3)]. These additionally weak dispersions are the replica of the valence bands shifted from $\overline{\Gamma}$ to $\overline{M}$, and conduction bands shifted from $\overline{M}$ to $\overline{\Gamma}$. For example, the V-shaped bands around the $\overline{M}$ point become centered around the $\overline{\Gamma}$ point after replication. The replica bands overlap with the original bands. The band replication suggests a 2 × 2 band folding and CDW order at 30 K. As shown in Fig. 3(g), a 2 × 2 reconstruction leads to BZ reconstruction in momentum space. The reconstruction vector can be described by three vectors $Q_1$, $Q_2$, and $Q_3$. Being the measurements are along the $\overline{M}$-$\overline{\Gamma}$-$\overline{M}$ direction, the observed folded bands are from $Q_1$ as shown by cyan double arrows in Figs. 3(b2) and 3(d2). Quantitative analysis of MDCs in Figs. 3(e) and 3(f) also confirmed this.

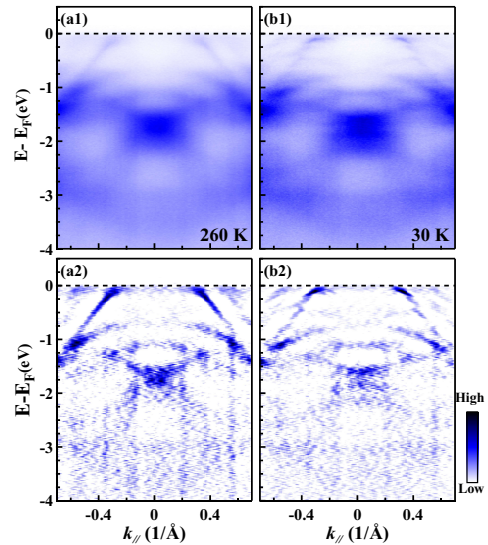

FIG. 2. **Broad range band structure of $Bi_2Rh_3Se_2$.** (a1) and (b1) Measured band structures along $\overline{K}$-$\overline{\Gamma}$-$\overline{K}$ direction taken with 30 eV photon energy at 260 K and 30 K, respectively. (a2) and (b2) Corresponding second-derivative images of (a1) and (b1), respectively.

In addition to electronic structure reconstruction, CDW transition often leads to band splitting and CDW gaps opening. From the low-temperature ARPES images [Figs. 3(b) and 3(d)], it can be seen that the replica bands interact with the original bands, resulting in the breaks in the original linear dispersion, as indicated by the white arrows [Figs. 3(b2), 3(b3), 3(d2), and 3(d3)]. The breaks in the dispersion at each intersection indicate the opening of the gap. There are a lot of bands crossing each other and generating a large number of intersections, which means that the back-folding induced "CDW gaps" open at a considerable number of different momentum and energy positions. Besides these gaps, real CDW gaps opening below the Fermi level associated to band splitting were observed at the high-symmetry points of the original BZ and the reconstructed BZ. As indicated by magenta arrows in Figs. 3(b1) and 3(b3), a large CDW gap $\Delta_1$ opens at $\overline{M}$ in the original BZ. The corresponding EDCs shown in Fig. 3(h) clearly shows the CDW gap $\Delta_1$ opening signatures. However, there is no quasiparticle peak at higher binding energy, so there is no way to obtain the energy gap size from EDCs. The size of CDW gap $\Delta_1$ obtained from the second derivative image Fig. 3(b3) is ∼ 120 meV. The red arrows in Figs. 3(b1) and 3(b3) indicate the CDW gaps opening at $\overline{M'}$ in the reconstructed BZ. The corresponding EDCs shown in Fig. 3(i) clearly shows the CDW gap $\Delta_2$ opening signatures with a size of ∼ 110 meV at 30 K determined by the separation between the two high-energy peaks. However, to exactly quantify the CDW gap size is a difficult task because the spectra are not smooth enough and there are no well-defined peaks at higher temperatures.

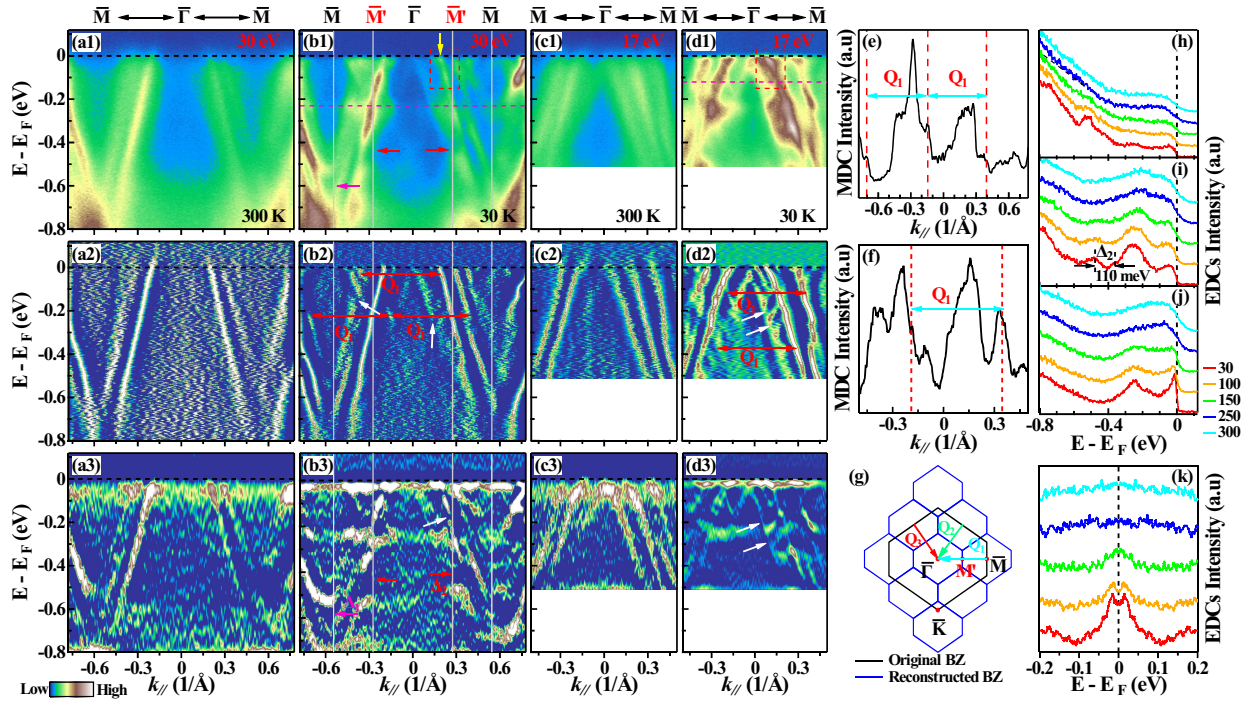

FIG. 3. **CDW-induced band reconstruction in $Bi_2Rh_3Se_2$.** (a1-d1)) Raw band maps of $Bi_2Rh_3Se_2$ along $\overline{M}$-$\overline{\Gamma}$-$\overline{M}$ direction taken at the labeled photon energies and temperatures. (a2-d2) Corresponding second-derivative images of (a1-d1) with respect to momentum. (a3-d3) Corresponding second-derivative images of (a1-d1) with respect to energy. (e) and (f) Momentum distribution curves (MDC) at the magenta horizontal dashed lines from the band maps in (b1) and (d1), respectively. The reconstruction wavevector $Q_1$ connects the original and folded bands. (g) The original (black lines) and 2 × 2 reconstructed (blue lines) BZs. The $\overline{\Gamma}$, $\overline{K}$, and $\overline{M}$ are the high-symmetry points of the original BZ. The $\overline{M'}$ is the high-symmetry point of the reconstructed BZ. The cyan, green, and red arrows indicate the possible electronic structure reconstruction wavevector $Q_1$, $Q_2$, and $Q_3$, respectively. The length of $Q_1$ is 0.545 $Å^{-1}$. The length of $Q_2$ and $Q_3$ is 0.470 $Å^{-1}$. (h) and (i) Temperature dependence of energy distribution curves (EDC) at $\overline{M}$ on the left, and $\overline{M'}$ on the right, respectively. Data were taken with 30 eV photons. The separation between EDC peaks represents the size of the CDW gap. (j) and (k) Temperature dependence of representative original and symmetrized EDCs. The location of the Fermi momentum is indicated by yellow arrow in (b1).

Generally, the Fermi surface nesting derived weak-coupling CDW should open a small gap on the Fermi surface. In $Bi_2Rh_3Se_2$, multiple Fermi surfaces are observed. According to the Fermi surface topology [Figs. 1(b) and 1(c)], if present, Fermi surface nesting is more likely to occur along the $\overline{\Gamma}$-$\overline{M}$ direction where parts of the Fermi surface sheets exhibit nearly parallel sections. Figure 3(j) shows temperature dependence of EDCs for a representative Fermi crossing as indicated by the yellow arrow in Fig. 3(b1). Strong temperature dependence of EDCs is observed. Upon cooling, a sharp "quasiparticle" peak develops near the Fermi level in the CDW state. To examine the temperature evolution of the energy gap, Fig. 3(k) show symmetrized EDCs at different temperatures. At 30 K, a small CDW gap of ∼ 15 meV is determined by half the separation between the two sharp peaks. Due to the limited experimental resolution, we have no way to determine the exact temperature at which the energy gap is closed. It is worth mentioning that the Fermi surface folder picture may also result in the opening of a small energy gap at Fermi surface. Our current data cannot confirm which one caused the opening of small energy gap at the Fermi surface.

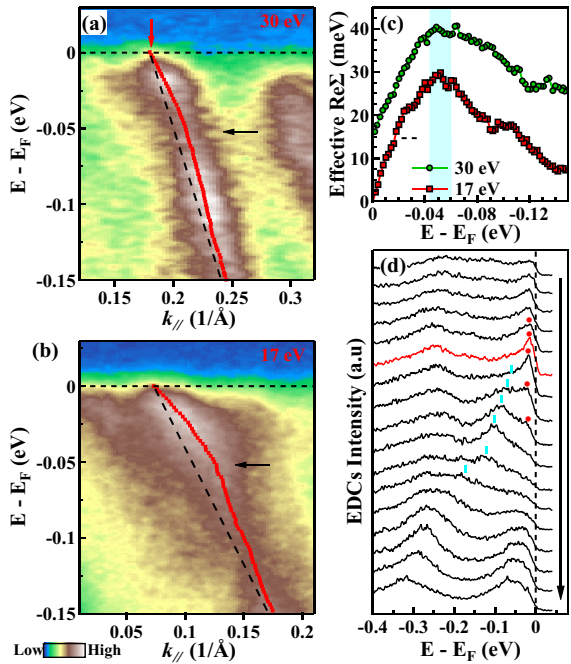

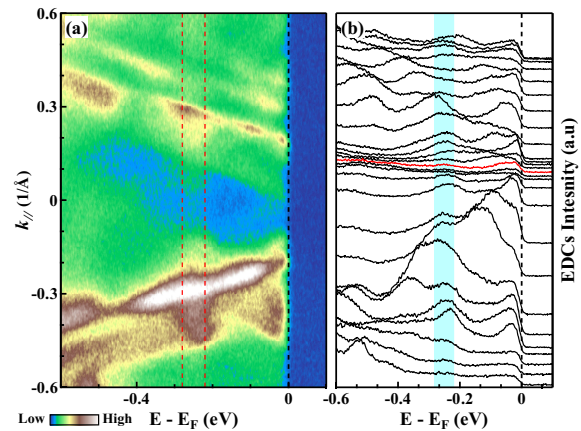

FIG. 4. ***e-ph* coupling in Bi$_2$Rh$_3$Se$_2$.** (**a**) and (**b**) Expanded band structures inside the red dashed rectangles in Figs. 3(b1) and 3(d1), respectively. The MDC fitted dispersions were shown in red. The black dashed lines represent the empirical bare band. (**c**) Effective real part of the self-energy ReΣ extracted from (a) and (b). (**d**) The corresponding EDCs of (a). Peak-dip-hump structure can be observed near the Fermi momentum, which is in red. The EDC peaks are marked by red dots, and the humps are marked by cyan bars. The black arrow indicate the direction of the momentum increase.

FIG. 5. **Flat band in Bi$_2$Rh$_3$Se$_2$.** (**a**) Band structure of Bi$_2$Rh$_3$Se$_2$ along $\overline{M}$-$\overline{\Gamma}$-$\overline{M}$ direction. (**b**) Corresponding EDCs of (a). The red curve represents EDC cross $\overline{\Gamma}$ point.

According to CDW formation theory, *e-ph* coupling is always in play. The *e-ph* coupling is also considered to be closely related to high-temperature superconductivity [26–28]. Obvious evidence of *e-ph* coupling can be observed in the band structures of Bi$_2$Rh$_3$Se$_2$. The zoom-in fragments as marked by red dashed rectangles in Figs. 3(b1) and 3(d1) corresponding to *e-ph* coupling areas are shown in Figs. 4(a) and 4(b). The prominent kink feature, a sign of *e-ph* coupling [12, 27–30], can be seen. The red squares are the dispersions obtained by fitting the MDCs at different binding energies. To obtain the kink energy quantitatively, linear black dashed lines are chosen as the empirical bare bands [28], and the effective real part of the electron self-energy (ReΣ) is obtained [Fig. 4(c)]. The effective ReΣ show pronounced peaks at ∼ 52 meV, which give rise to the kink in dispersion seen here. The EDCs corresponding to Fig. 4(a) are shown in Fig. 4(d). A peak-dip-hump feature can be found near the Fermi momentum (red curve), which is also a sign of *e-ph* coupling [27, 29]. Therefore, we believe that *e-ph* coupling may play an important role in the CDW phase transition of Bi$_2$Rh$_3$Se$_2$.

After studying the electronic structure associated with CDW order in Bi$_2$Rh$_3$Se$_2$, we'd like to introduce another interesting find, the flat band, which can be observed in the second derivative images Figs. 3(b3) and 3(d3). Flat bands are present in various materials, such as transition-metal dichalcogenides [31], 2D electron liquid [32], and strongly corrected materials such as Bi$_2$Sr$_2$CaCuO$_{8+\delta}$ [33], twisted bi-layer graphene [34], and kagome lattice CsV$_3$Sb$_5$ [35]. Figure 5(a) replots the ARPES image of Bi$_2$Rh$_3$Se$_2$ shown in Fig. 3(b1) aims for better visualization of the flat band. A flat band with binding energy around 0.25 eV can be seen in the entire momentum space. This feature can also be seen from the EDCs shown in Fig. 5(b). Considering the large energy scale of the flat band and the lack of its replica bands, it is unlikely to originate from impurity band emissions [36], *e-ph* coupling [32, 37], polaron effect [38], or nondispersive excitation [35]. Therefore, we consider that the flat band may originate from hybridization spreading [39].

In summary, we have studied the electronic structure of ternary superconductor Bi$_2$Rh$_3$Se$_2$ using ARPES. Our data demonstrate that the high-temperature phase transition at 240 K is a second-order CDW phase transition. Our study provides clear evidence of the emergence of a 2 × 2 CDW order at low temperature, accompanied by CDW induced electronic structure reconstruction, the associated band folding, band splitting, and multiple CDW gaps opening at and away from the Fermi level. We also have observed the kink and peak-dip-hump feature, demonstrating the presence of *e-ph* coupling. We believe that *e-ph* coupling may play an important role in the formation of CDW order. Furthermore, a flat band located at ∼ 0.25 eV below Fermi level was observed. Our findings will help us to gain an in-depth understanding of the origin of the CDW order and its possible relationship with superconductivity in the Bi$_2$Rh$_3$Se$_2$ superconductor.

This work was supported by the National Natural Science Foundation of China (Grant No. 12074436), and the Innovation-driven Plan in Central South University (Grant No. 2016CXS032).

---

* Corresponding author: jqmeng@csu.edu.cn

# Supplemental Material:

# Charge Density Wave order and electron-phonon coupling in ternary superconductor $Bi_2Rh_3Se_2$

Zi-Teng Liu,[1] Chen Zhang,[1] Qi-Yi Wu,[1] Hao Liu,[1] Bo Chen,[1] Zhi-Bo Yin,[1] Sheng-Tao Cui,[2] Zhe Sun,[2] Shuang-Xing Zhu,[1] Jiao-Jiao Song,[1] Yin-Zou Zhao,[1] Hong-Yi Zhang,[1] Xue-Qing Ye,[1] Fan-Ying Wu,[1] Shu-Yu Liu,[1] Xiao-Fang Tang,[3] Ya-Hua Yuan,[1] Yun-Peng Wang,[1] Jun He,[1] Hai-Yun Liu,[4] Yu-Xia Duan,[1] and Jian-Qiao Meng [1, †]

[1]*School of Physics and Electronics, Central South University, Changsha 410083, Hunan, China*
[2]*National Synchrotron Radiation Laboratory, University of Science and Technology of China, Hefei 230029, Anhui, China*
[3]*Department of Physics and Electronic Science, Hunan University of Science and Technology, Xiangtan 411201, Hunan, China*
[4]*Beijing Academy of Quantum Information Sciences, Beijing 100085, China*

**The supplemental materials to 'Charge Density Wave order and electron-phonon coupling in ternary superconductor $Bi_2Rh_3Se_2$' contains 'Temperature evolution of the folded bands', and 'Determination of CDW phase transition temperature'.**

## 1. Temperature evolution of the folded bands

A detailed view of the folded bands taken along $\overline{\text{M}}$-$\overline{\Gamma}$-$\overline{\text{M}}$ direction is shown in Fig. S1. Above phase transition temperature, the electronic structure is dominated by linear dispersions, indicating a normal metallic state. At temperatures well below the phase transition, the original linear dispersions are overlaid by replica bands, resulting in a very complex band structure. The replica bands vary significantly with temperature. The intensity of replica bands diminishes as temperature increases and becomes indiscernible around 200 K and above. However, due to the complexity of the band structures, it is impossible to obtain an accurate phase transition temperature from this set of data.

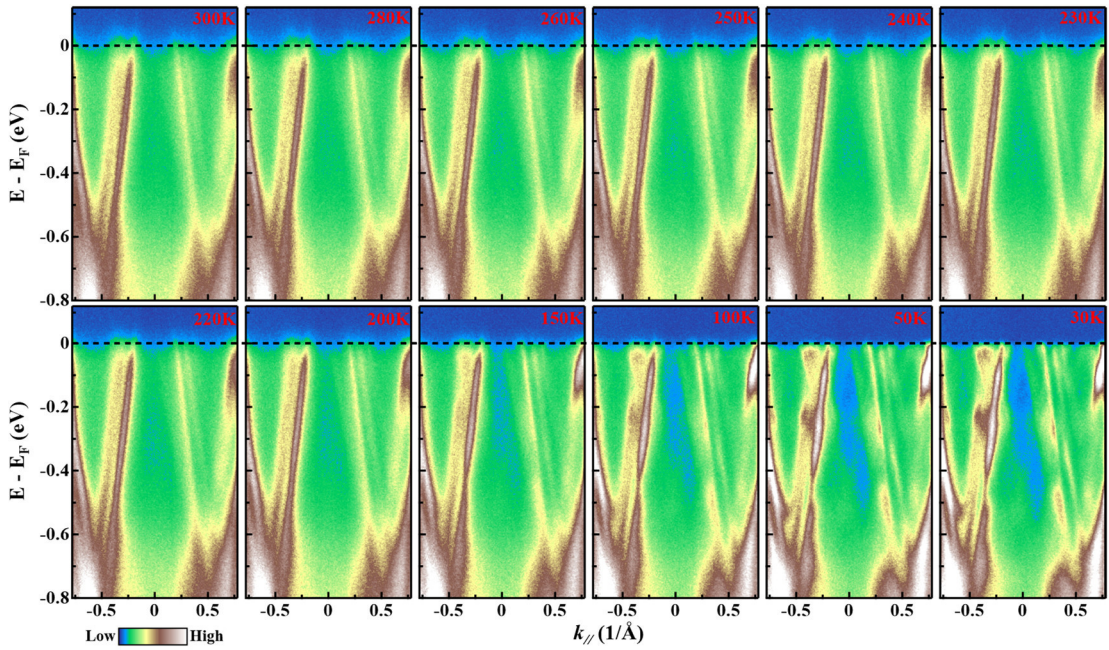

**Fig. S1** Temperature dependence of the folded bands. Measurements were taken along $\overline{\text{M}}$-$\overline{\Gamma}$-$\overline{\text{M}}$ direction with 30 eV phonons at labeled temperature.

## 2. Determination of CDW phase transition temperature

To determine the phase transition temperature by ARPES, detailed temperature measurements were carried out along the $\overline{\text{K}}$-$\overline{\Gamma}$-$\overline{\text{K}}$ direction. This direction was chosen because the band structure in this direction is simpler. Fig. S2 shows the original maps measured with 17 eV photons at labeled temperatures. As can be seen from the figure, energy bands have temperature-dependent behaviors similar to those shown in Fig. S1. We quantitatively studied the temperature dependence of spectra intensity at the location of the replica bands appearing, which is marked by black dashed rectangle on the image of 30 K. Fig. S3 displays the integrated intensity of the interest regions. As the temperature decreases, the intensity of the replica bands begins to increase significantly around the phase transition temperature $T_S$. The integrated intensity, black squares, representing the replica bands can be well fitted by an empirical mean-field equation describing the second-order phase transition, as follows

$$I(T) \propto \tanh^2(\alpha\sqrt{\frac{T_C}{T}-1})\Theta(T_C - T)$$

where α is a fitting parameter, $T_C$ is the phase transition temperature, and Θ is the Heaviside step function. The fitting result is shown by the green line in Fig. S3, giving a transition temperature $T_{CDW}$ = 244 ±4 K. The transition temperature obtained in our study is consistent with its reported high temperature phase transition temperature. A very good fit also means that this is a second-order CDW phase transition.

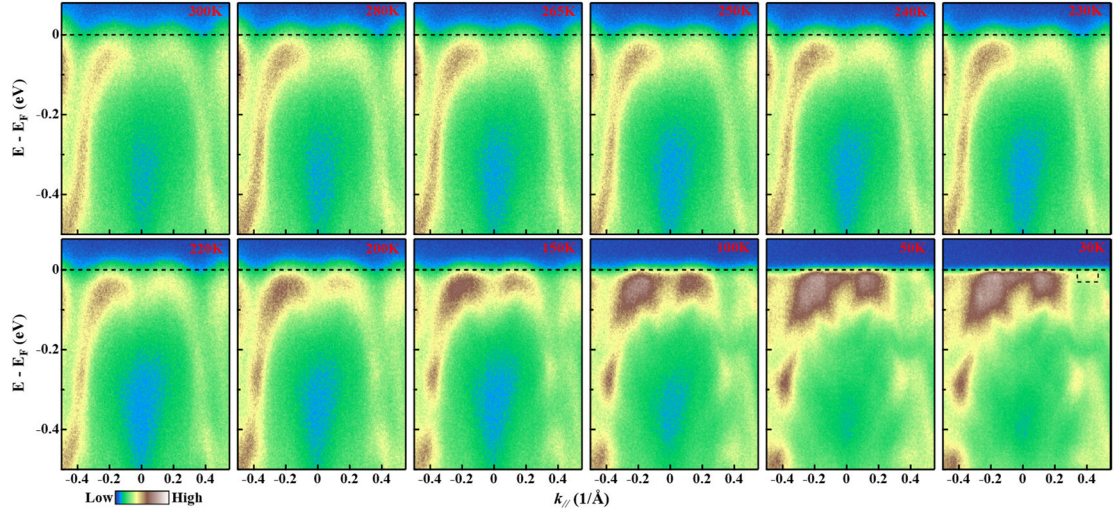

**Fig. S2** Temperature dependence of the folded bands. Measurements were taken along $\bar{K}$-$\bar{\Gamma}$-$\bar{K}$ direction with 17 eV phonons at labeled temperature.

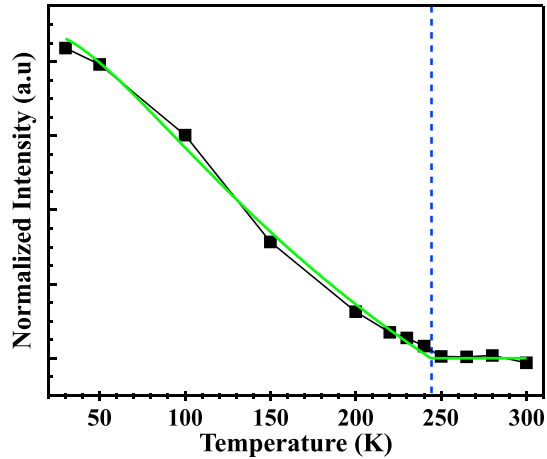

**Fig. S3** Temperature-dependent integrated ARPES intensities of the interest region. The green line shows the fitting curve.